\documentclass[12pt]{article}
\usepackage{amssymb}
\usepackage{amsmath}
\usepackage{graphicx}
\usepackage{soul}
\usepackage{xcolor}
\begin{document}
\makeatletter
\@addtoreset{equation}{section}
\makeatother
\renewcommand{\theequation}{\thesection.\arabic{equation}}
%
%

\title{Tachyon with an inverse power-law potential in a braneworld cosmology}

\author{
Neven Bili\'c$^1$\thanks{bilic@irb.hr}, Silvije Domazet$^1$\thanks{sdomazet@irb.hr},
and Goran S.\ Djordjevic$^2$\thanks{gorandj@junis.ni.ac.rs}
   \\
   $^1$Division of Theoretical Physics, Rudjer Bo\v{s}kovi\'{c} Institute,\\ Zagreb,
   Croatia\\
$^2$Department of Physics,
University of Nis,  Serbia\\
}

\maketitle

\begin{abstract}
We study  a tachyon cosmological model  based on the dynamics of a 3-brane
in the bulk of the second Randall-Sundrum  model extended to more general warp functions.
A well known prototype of such a generalization is the bulk with a selfinteracting scalar field.
As a consequence of a generalized bulk geometry the cosmology on the observer brane is modified 
by the scale dependent four-dimensional gravitational constant. 
In particular we study a power law warp factor which generates
an  inverse power-law  potential 
$V\propto \varphi^{-n}$ of the tachyon field $\varphi$.
We find a critical power $n_{\rm cr}$ that divides two
subclasses with distinct asymptotic behaviors: a dust universe  for $n>n_{\rm cr}$ and a quasi de Sitter universe for 
$0<n<n_{\rm cr}$. 
  \end{abstract}


\section{Introduction}	


Braneworld cosmology is based on the  scenario in which matter is confined 
on a brane moving in the higher dimensional bulk
with only gravity allowed to propagate in the bulk  \cite{arkani,antoniadis}.
Among many braneworld models a particularly important are the two versions of the Randall-Sundrum (RS) model.
 The first RS model (RSI) \cite{randall1} was
originally proposed as a solution to the hierarchy problem in particle physics whereas
the second RS model (RSII) \cite{randall2} renders 
a mechanism for localizing gravity on the 3+1 dimensional universe embedded in
a 4+1 spacetime without compactification of the extra dimension.

Immediately after the papers  \cite{randall1,randall2} appeared, it was realized that the RS model
is immersed in a wider framework of the so called AdS/CFT correspondence 
\cite{gubser2} (for a recent retrospect see appendix of Ref.\ \cite{bilic1}).
At the same time it was realized that the RS model, as well as similar braneworld models, 
may have interesting cosmological
implications \cite{binetruy,flanagan}. In particular, owing to the presence of an extra dimension  
and the bulk cosmological constant related to the brane tension, the usual 
Friedmann equations are modified
\cite{binetruy}
so the model can have predictions different from the standard cosmology
and is therefore subject to cosmological tests \cite{godlowski}.

The RS model, originally proposed with a pure 4+1-dimensional anti-de Sitter (AdS$_5$) bulk, 
can be extended to include
matter in the bulk.
A massive scalar field in the bulk was first introduced by Goldberger and Wise \cite{goldberger} 
to stabilize the brane separation in RSI.
The RS model with a minimally coupled scalar field in the bulk, referred to as the {\em thick brane} model,
has been constructed for maximally symmetric solutions on the brane 
 \cite{kobayashi}.
It has been demonstrated that the bulk scalar potential and the corresponding
brane potential can be 
 derived from a superpotential in which case the solution of the static vacuum geometry reduces to a set
of first-order BPS-like equations \cite{dewolfe}.

A noncanonical scalar field in the bulk with bulk tachyon Lagrangian has been considered 
in Ref.\  \cite{german1} where  a thick braneworld 
has been investigated in the cosmological context. In particular, 
the stability under tensor perturbations and  gravity localization has been demonstrated.
In a subsequent paper \cite{german2} the stability under scalar perturbation has been studied for a
braneworld with maximally symmetric geometry.
Models where matter in the bulk contains a non-minimally coupled selfinteracting scalar have also been studied.
Some interesting features of these models can be found in Refs.\ \cite{Bogdanos:2006dt,HerreraAguilar:2011jm} 
and references therein.

In this paper we study an 
RSII-type braneworld cosmology extended to more general warp factors.
As an application, we study in particular 
a braneworld  scenario based on this 
extended RSII with an effective  tachyon field on 
the brane\footnote{Note that our tachyon field is located on the observer brane
in contrast to the bulk tachyon of
Refs.\  \cite{german1,german2}.}.
What distinguishes the tachyon  from the canonical 
scalar field is  
the Lagrangian of the Dirac-Born-Infeld (DBI) form \cite{sen}:
\begin{equation}
 \mathcal{L}=V(\varphi) \sqrt{1-g^{\mu\nu}\varphi_{,\mu}\varphi_{,\nu}} .
\end{equation}
A similar Lagrangian  appears in the so called DBI
inflation models \cite{shandera}. In these models the inflation
is driven by the motion of a D3-brane in a warped throat
region of a compact space and the 
DBI field corresponds to the position of the D3-brane.

As shown by Abramo and Finelly \cite{abramo} in a standard cosmological scenario, 
for the class of tachyon models 
with inverse power-law potentials $V(\varphi) \propto \varphi^{-n}$,
the power $n=2$ divides two subclasses with distinct behaviors in the asymptotic regimes.
For $n<2$ 
in the limit $\varphi \rightarrow \infty$, the pressure 
 $p\rightarrow -1$ and the universe 
behaves as quasi-de Sitter.
For $n>2$, for large $\varphi$ the pressure $p\rightarrow 0^{-}$ very quickly 
yielding asymptotically a cold dark matter (CDM) domination.
In the context of tachyon inflation,
in both cases after the inflationary epoch the tachyon will remain a dominant component  
unless at the end of inflation, it
  decayed into inhomogeneous fluctuations and
other particles. This period, known as reheating
\cite{kofman2,kofman3,bassett},  links
the inflationary epoch with the subsequent thermalized
radiation era. 

A simple tachyon model can be realized in the framework of RSII.
 The original RSII consists of two 3-branes in 
the AdS$_5$ bulk 
with line element
\begin{equation}
ds^2_{(5)}=G_{ab} dX^a dX^b=e^{-2|y|/\ell} \eta_{\mu\nu}dx^\mu  dx^\nu 
  -dy^2 ,
 \label{eq3000}
\end{equation}
with observer's brane placed at $y=0$ and the negative tension brane pushed of to $y=\infty$.
It may be easily shown \cite{bilic} that one additional 3-brane moving in the AdS$_5$ bulk
behaves effectively as a tachyon with a potential
$V(\varphi) \propto \varphi^{-4}$ and hence 
drives a dark matter attractor.
A more general tachyon potential could be obtained from
 more general bulk geometry. 
This can be achieved if one  assumes a presence of matter in the bulk, e.g., 
in the form of a selfinteracting scalar field.
The bulk scalar would change the bulk 
geometry depending on the scalar field potential. 
In addition, the braneworld cosmology would differ from that of the original RSII model.  

A straightforward approach would be to start from a given bulk field potential and
derive the bulk geometry which would, in turn, yield a tachyon potential
of the effective tachyon field induced by the dynamical 3-brane.
A more empirical approach would be to  
go the other way round: starting from a given, phenomenologically interesting tachyon potential
one would fix the warp factor and one could, in principle, 
construct the bulk scalar-field selfinteraction potential. 
In this paper we will  start from a warp factor of a general form
and  use the tachyon  model to study the cosmology on the brane.
In particular, we will study the effects of the warp factor of the power-law form which may be linked to
the exponential superpotential of the bulk field. 
With this warp the tachyon will have an inverse power-law potential. 
We will analyze this type of tachyon  
potentials in four different cosmological scenarios: the standard tachyon cosmology, low  
and high density regimes of our branenworld model, and high density Gauss-Bonnet braneworld cosmology. 

The remainder of the paper is organized as follows. 
In the next section we introduce  the extended  RSII 
and derive the corresponding braneworld cosmology. In Sec. \ref{dynamical}
we introduce the tachyon as a dynamical brane and derive the field equations
in a covariant Hamiltonian formalism.
The  asymptotic solutions to the field equations for
the inverse power law potential are presented in Sec.\ \ref{solutions}. 
In Sec. \ref{conclude} we give the concluding remarks.
Finally, in Appendix \ref{scalar} we outline the derivation of
the generalized RSII model with a scalar field in the bulk.

\section{Braneworld cosmology}
\label{braneworld}

Our curvature conventions are as follows:
$R^{a}{}_{bcd} = \partial_c \Gamma_{db}^a - 
\partial_d \Gamma_{cb}^a + \Gamma_{db}^e \Gamma_{ce}^a  - \Gamma_{cb}^e \Gamma_{de}^a$ 
and $R_{ab} = R^s{}_{asb}$, 
so Einstein's equations are $R_{ab} - \frac{1}{2}R G_{ab} = +8\pi G T_{ab}.$

Here, we 
derive the braneworld cosmology assuming that the bulk spacetime is 
given by the metric
\begin{equation}
ds^2_{(5)}=G_{ab} dX^a dX^b=\psi(y)^2 \eta_{\mu\nu}dx^\mu  dx^\nu 
  -dy^2 ,
 \label{eq3006}
\end{equation}
and the cosmology is determined by the motion of the brane.
It will be sometimes advantageous to work  in conformal coordinates  with line element
\begin{equation}
ds^2_{(5)}=\frac{1}{\chi(z)^2}( g_{\mu\nu}dx^\mu  dx^\nu 
  -dz^2).
 \label{eq4112}
\end{equation}
The two metrics are related by the coordinate transformation 
\begin{equation}
dz= \frac{dy}{\psi(y)} 
 \label{eq3011}
\end{equation}
and
\begin{equation}
\chi(z)=\frac{1}{{\psi(y(z))}}.
 \label{eq3010}
\end{equation}
The observer brane is placed at $y=y_{\rm br}$ ($z=z_{\rm br}$) and,
as in the original  RSII model, we  assume
the $Z_2$ orbifold symmetry $y-y_{\rm br}\leftrightarrow y_{\rm br}-y$ ($z\leftrightarrow z_{\rm br}^2/z$),  so
the region $-\infty<y\leq y_{\rm br}$ ($0<z\leq z_{\rm br}$) is identified with 
$ y_{\rm br}\leq y <\infty$ ($z_{\rm br} \leq z <\infty$). 

Next, we assume
that  observer's brane  has additional matter
represented by the Lagrangian $\mathcal{L}$,
i.e., the brane action is
\begin{equation} 
S_{\rm br}[h] =\int_\Sigma d^{4}x\sqrt{-h} (-\lambda(y) + \mathcal{L}[h]) ,
\end{equation} 
with a $y$-dependent  brane tension $\lambda$.
Then, we allow the brane  
to move in the bulk along the fifth coordinate $y$.
In other words, the brane hypersurface $\Sigma$ 
is time dependent and may be
defined by 
\begin{equation}
\psi(y)^2-a(t)^2=0 ,
\label{eq027}
\end{equation}
where $a=a(t)$ is an arbitrary function.
The normal to $\Sigma$ is then given by
\begin{equation}
n_\mu \propto\partial_\mu (\psi(y)-a(t))=(-\dot{a},0,0,0,\psi')
\label{eq013}
\end{equation}
and, using the normalization $g^{\mu\nu}n_\mu n_\nu=-1$, one finds the nonvanishing components
\begin{equation}
n_t =\frac{\dot{a}}{\psi'}\left(1 -\frac{\dot{a}^2}{\psi^2\psi^{\prime 2}}\right)^{-1/2}, 
\label{eq014}
\end{equation}
\begin{equation}
n_y=-\left(1 -\frac{\dot{a}^2}{\psi^2\psi^{\prime 2}}\right)^{-1/2} . 
\label{eq0141}
\end{equation}
Using this,  we find the induced line element on the brane 
\begin{equation}
ds_{\rm ind}^2=(G_{ab} +n_an_b) dx^adx^b=n(t)^2dt^2 -a(t)^2 d{\bf x}^2 ,
\label{eq028}
\end{equation}
where 
\begin{equation}
n^2 =a^2-\frac{\dot{a}^2}{\psi^{\prime 2}} .
\label{eq016}
\end{equation} 
It is understood that the argument $y$ of  $\psi'(y)$
is implicitly time dependent through (\ref{eq027}).
The Friedmann equations on the brane follow directly from
the junction conditions \cite{israel}
\begin{equation}
\left[ \left[ K^{\mu}_{\nu} - \delta^{\mu}_{\nu}  K_{\alpha}^{\alpha} \right] \right]
= 8 \pi G_5  (\lambda(y)\delta^{\mu}_{\nu}+ T^{\mu}_{\nu}) ,
\label{eq099}
\end{equation}
where $ K^{\mu}_{\nu}$ is the pullback of the extrinsic 
curvature tensor
defined in Appendix (Eq.\ (\ref{eq109})).
The energy momentum tensor $T^{\mu}_{\nu}={\rm diag} (\rho,-p,-p,-p)$ corresponds to the Lagrangian
$\mathcal{L}$. From (\ref{eq099}) together with
(\ref{eq014})-(\ref{eq016}) we obtain
\begin{equation}
\frac{(\partial_t a)^2}{n^2 a^2} +\frac{\psi^{\prime 2}}{\psi^2}=\left(\frac{4\pi G_5}{3}\right)^2
(\lambda+\rho)^2 .
\label{eq020}
\end{equation}
The first term on the left-hand side of (\ref{eq020}) is the square of 
the Hubble expansion rate for the metric (\ref{eq028}) on the brane
\begin{equation}
 H^2=\frac{(\partial_t a)^2}{n^2 a^2}.
\label{eq021}
\end{equation}
Then, the first Friedmann equation takes  the form
\begin{equation}
H^2= \frac{8\pi G_{\rm N}}{3}\left(\frac{4\pi G_5}{3k}\lambda\right) \rho+
\left(\frac{4\pi G_5}{3}\right)^2\rho^2 +
\left(\frac{4\pi G_5}{3}\lambda\right)^2 -\frac{\psi^{\prime 2}}{\psi^2},
\label{eq022}
\end{equation}
where $G_{\rm N}$  is the four-dimensional Newton constant  and $k$ is a mass scale related to $G_5$,
\begin{equation}
k= \frac{G_{\rm N}}{G_5} .
\label{eq023}
\end{equation}
Generally, $\lambda$, $\psi$ and $\psi'$ are implicit functions of $a(t)$ through their dependence
on $y$ which in turn is a function of $a$ via (\ref{eq027}). 
For a pure AdS$_5$ bulk with curvature radius $\ell=1/k$ we have  
$ \psi^{\prime}/\psi=-k$,  $k=1/\ell$, $\lambda$ is constant and with the RSII fine tuning condition 
 we recover the usual RSII expressions (see, e.g., Ref. \cite{bilic1}).
 Henceforth we will assume  $|\psi^{\prime}/\psi|\lesssim k$ and the average of $|\psi^{\prime}/\psi|$
will basically represent a warp compactification scale.

A modified Friedmann equation  similar to (\ref{eq022})  
was derived  by P.~Brax, C.~van de Bruck and A.~C.~Davis \cite{brax2}
for a braneworld with a scalar field $\Phi$ in the bulk
with a time dependent geometry.
The full expression in our notation  reads  
\begin{equation}
H^2=\frac{4\pi G_5}{9}W\rho+\left( \frac{8\pi G_5}{3}\rho\right)^2-\frac{4\pi G_5}{9a^4}
\int d\tau\frac{dW}{d\tau} a^4\rho -\frac{1}{6a^4}\int d\tau\frac{da^4}{d\tau}
\left(\frac12 (\partial_\tau\Phi)^2-U_{\rm eff}\right), 
 \label{eq315}
\end{equation}
where $\tau$ is the synchronous time and  $W=W(\Phi)$ is the superpotential. The 
effective potential $U_{\rm eff}=U_{\rm eff}(\Phi)$
is defined as
\begin{equation}
U_{\rm eff}=\frac{W^2}{6}-\frac18 \left(\frac{dW}{d\Phi}\right)^2+U
 \label{eq316}
\end{equation}
and $U=U(\Phi)$ is the bulk-field potential. 
It is understood that $\Phi$ and its derivative $\partial_\tau\Phi$  
are functions of $\tau$ only.
The contribution of the last two terms in (\ref{eq315}) is  referred to as the {\em retarded effect}
\cite{brax2}.
In deriving (\ref{eq315}) the contribution of dark radiation has not been taken into account.

 Eq.\ (\ref{eq315}) reduces to a simpler equation similar to our   
(\ref{eq022}) if one assumes 
that the bulk scalar field is evolving much slower than the scale factor.
On this assumption we have
\begin{equation}
\frac{1}{W}\frac{dW}{d\tau} \ll \frac{1}{a^4\rho}\frac{da^4\rho}{d\tau}
 \label{eq317}
\end{equation}
and 
\begin{equation}
\frac{da^4}{d\tau}
\left(\frac12 (\partial_\tau\Phi)^2-U_{\rm eff}\right)\simeq 
\frac{d}{d\tau}
\left(\frac{a^4}{2} (\partial_\tau\Phi)^2-a^4U_{\rm eff}\right), 
 \label{eq333}
\end{equation}
so the third term on the righthand side of (\ref{eq315}) can be neglected compared with
the first term and 
the integration in the last term is trivially performed.
 The constant of integration can be set to zero as it
 would only contribute to 
a dark radiation term of the form $1/a^4$ which has been ignored anyway. Hence, up to 
a dark radiation term, Eq. (\ref{eq315}) reduces to
\begin{equation}
H^2=\frac{4\pi G_5}{9}W\rho+\left( \frac{8\pi G_5}{3}\rho\right)^2+\frac{1}{6}
\left(U_{\rm eff}-\frac12 (\partial_\tau\Phi)^2\right).
 \label{eq318}
\end{equation}

Now we show explicitly that our equation (\ref{eq022}) is equivalent to (\ref{eq318}).
First we identify the superpotential and bulk potential 
\begin{equation}
W(\Phi)\equiv 8\pi G_5 \lambda(y),
 \label{eq320}
\end{equation}
\begin{equation}
U(\Phi)\equiv\frac12 \Phi^{\prime 2} -6\frac{\psi^{\prime 2}}{\psi^2},
 \label{eq330}
\end{equation}
where $\Phi=\Phi(y)$ is  yet unspecified function of $y$.
Next, using 
Eqs.\ (\ref{eq316}), (\ref{eq320}), and (\ref{eq330}) as the defining equations for $U_{\rm eff}$,
we can express (\ref{eq022})
in the form
\begin{equation}
H^2=\frac{4\pi G_5}{9}W\rho+\left( \frac{8\pi G_5}{3}\rho\right)^2+\frac{1}{6}U_{\rm eff}-
\frac{1}{12}\left(\Phi^{\prime 2}-\frac14\left(\frac{\partial W}{\partial\Phi}\right)^2\right).
 \label{eq331}
\end{equation}
Finally, by demanding that the function $\Phi$ satisfies
\begin{equation}
\Phi^{\prime 2}-\frac14\left(\frac{\partial W}{\partial\Phi}\right)^2=\left(\partial_\tau\Phi\right)^2,
 \label{eq332}
\end{equation}
our equation  (\ref{eq022}) takes the form identical to (\ref{eq318}).
Hence, we have demonstrated that equations (\ref{eq022}) and (\ref{eq318}) are equivalent.

By manipulating Eq. (\ref{eq332}) with the help of (\ref{eq013}), (\ref{eq016}), and (\ref{eq021})
we find that $\Phi$ satisfies another equation equivalent to (\ref{eq332}):  
\begin{equation}
\Phi^{\prime}=\frac12 \frac{\partial W}{\partial\Phi}\left(1-H^2\frac{\psi^2}{\psi^{\prime 2}}\right)^{-1/2}.
 \label{eq319}
\end{equation}
Given the functions $\lambda(y)$ and $\psi(y)$,  equations (\ref{eq320}), (\ref{eq330}), and 
(\ref{eq319}) together with (\ref{eq022}) determine parametrically  $W$ and $U$ as functions of $\Phi$.
Note that Eqs.\ (\ref{eq319}) 
is consistent with the first equation in (\ref{eq4007}) in the static limit $H\rightarrow 0$.

In this way, the scale dependence of the brane tension $\lambda(y)$ which has not been specified yet,
can be attributed to  a scalar field dynamics in the bulk.
Due to the retarded effect,  equation (\ref{eq315}), even in its reduced form (\ref{eq022}),
is rather complicated.
However, we can simplify our braneworld cosmology by assuming that the contribution of 
the retarded effects in (\ref{eq022}) are negligible compared to $H^2$, i.e., we will assume
\begin{equation}
\left(\frac{4\pi G_5}{3}\lambda\right)^2\approx\frac{\psi^{\prime 2}}{\psi^2}.
 \label{eq335}
\end{equation}
This assumption is motivated by the junction condition (\ref{eq303})
(as a generalization of the fine tuning condition of RSII)
which is exact in the static case.
Then our approximated braneworld cosmology is defined by the Friedmann equation
\begin{equation}
H^2= \frac{8\pi G}{3} \rho+\left(\frac{4\pi G_{\rm N}}{3k}\rho\right)^2 .
\label{eq005}
\end{equation}
Here $G=G(a)$ is a scale dependent effective gravitational constant defined as 
\begin{equation}
 G(a) = \frac{G_{\rm N}}{k}\left.\frac{d\chi}{dz}\right|_{z=\chi^{-1}(1/a)} ,
 \label{eq007}
\end{equation}
where $\chi^{-1}$ denotes the inverse  function  of $\chi$.
The second Friedmann equation is easily obtained by combining the time derivative of (\ref{eq005})
with energy-momentum conservation yielding
\begin{equation}
\dot{H}= -\left(4\pi G+3\left(\frac{4\pi G_{\rm N}}{3k}\right)^2 \rho\right) (p+\rho)+
\frac{4\pi}{3}\frac{dG}{da}a\rho .
\label{eq4205}
\end{equation}

As a promising  future research topic
it would be of interest to extend our approach along the lines of Ref.\
\cite{bernardini}
where the warp factor was allowed to depend on the radial braneworld coordinate $r$
in addition to the usual $y$ and $t$ dependence.
A natural extension worth of investigating would be including  $r$
dependence of the bulk scalar field in addition to its
 $y$ dependence.

In the following we shall  abbreviate by BWC 
the braneworld cosmology described by (\ref{eq005}) and (\ref{eq4205}).
Thus,  the  Friedmann equations of  BWC differ from those of the original RSII  
in the scale dependence of the effective gravitational constant $G$ and
in one additional term in the second equation that depends on the derivative
of $G$. This term  will be suppressed if the scale dependence of $G$ is weak.

Equation (\ref{eq007}) imposes certain restrictions on the function $\chi$.
First, we need $G(a)$ to be positive which restricts  
$\chi(z)$ to the class of monotonously increasing functions of $z$ and,
as a consequence of (\ref{eq3011}), $\psi$ must be a monotonously decreasing function of $y$.
Second, the variation of $G(a)$ is constrained by the big bang nucleosynthesis \cite{accetta} 
and other cosmological and astrophysical
observations \cite{uzan}. The observations are roughly consistent with the constraint
\begin{equation}
 \left|\frac{\dot{G}}{G}\right|_{\rm today} \lesssim 10^{-12} {\rm yr}^{-1}
\end{equation}
which, in turn, implies
\begin{equation}
\left|\frac{a}{G}\frac{dG}{da}\right|_{\rm today}=\frac{1}{H_0}
\left|\frac{\dot{G}}{G}\right|_{\rm today}
\lesssim 1.43\times10^{-2},
\label{eq006}
\end{equation}
where we have used the value $H_0^{-1}=14.3$ Gyr corresponding to
the Planck 2015 estimate of the Hubble constant $H_0=68$ km/s/Mpc \cite{planck2015}. 
Using the definition (\ref{eq007}), we find a relation
\begin{equation}
\frac{a}{G}\frac{dG}{da}=-\left. 
\frac{\chi \chi_{,zz}}{\chi_{,z}^2}\right|_{z=\chi^{-1}(1/a)}.
\label{eq009}
\end{equation}
Thus, equation (\ref{eq006}) imposes a constraint also on $\chi(z)$.
For our purpose the metric of the power law form 
$\chi\propto z^{n/4}$ is of particular interest.
In this case 
Eq. (\ref{eq009}) reads 
\begin{equation}
\frac{a}{G}\frac{dG}{da}=-\frac{n-4}{n}
\label{eq314}
\end{equation}
and  Eq. (\ref{eq006}) imposes a constraint on the power
\begin{equation}
|n-4| \lesssim 0.057,
\label{eq1004}
\end{equation}
where the central value  $n=4$ corresponds to the original RSII setup with constant $G$.

As demonstrated in Appendix \ref{scalar}, the power-law warp $\chi\propto z^{n/4}$ with $n\geq 4$ can be 
attributed to a selfinteracting scalar field $\Phi$ in the bulk with
exponential superpotential $W\propto \exp \gamma \Phi$,
Then, the constraint (\ref{eq1004}) implies  a constraint on the
parameter $\gamma$ (see also Ref. \cite{amarilla})
\begin{equation}
\gamma^2  \lesssim 0.477\times 10^{-2} ,
\label{eq1008}
\end{equation}
which is less stringent than the order of magnitude estimate \cite{davis}
$\gamma\lesssim 0.01$ based on the solar system bounds on the Edington parameter \cite{bertotti}.

The constraint  (\ref{eq006}) yielding (\ref{eq1004}) and (\ref{eq1008}) 
is obtained from  astrophysical and cosmological
bounds on variation of $G$ for the period  between BBN  
(corresponding roughly to the scales $a\simeq 10^{-9}$) and today.
It is conceivable that the bounds on variations of $G$ with $a$  are less restrictive
for the early cosmology prior to BBN.

\section{Dynamical brane as a tachyon}
\label{dynamical}

In this section we introduce a tachyon via the dynamical brane.
In addition to the observer brane at $y=y_{\rm br}$ we place a non BPS positive tension 
brane at $y>y_{\rm br}$ in the bulk with  
metric (\ref{eq3006}). Our setup is similar to that of Lykken and Randall \cite{lykken}. 

The action of the 3+1 dimensional brane in the five dimensional bulk is equivalent 
to the Dirac-Born-Infeld description of a Nambu-Goto 3-brane.
\cite{bordemann,jackiw}.
Consider a 3-brane 
 moving in the 4+1-dimensional bulk spacetime 
  with coordinates $X^{a}$,
$a=0,1,2,3,4$. The points on the brane are parameterized by
$X^a (x^{\mu})$, $\mu=0,1,2,3$, where $x^{\mu}$
are the coordinates on the brane.
The brane action is given by  
\begin{equation}
S_{\rm br}= - \sigma
\int d^4x\, \sqrt{-\det (g^{(\rm ind)})}  \, , 
\label{eq0001}
\end{equation}
where $\sigma$ is the brane tension and  $g_{\mu\nu}^{(\rm ind)}$ is the induced metric
or the ``pull back" of the bulk space-time metric 
$G_{ab}$ to the brane,
\begin{equation}
g^{(\rm ind)}_{\mu\nu}=G_{ab}
\frac{\partial X^a}{\partial x^\mu}
\frac{\partial X^b}{\partial x^\nu} \, .
\label{eq0002}
\end{equation}
Taking the Gaussian normal parameterization 
 $X^a(x^\mu)=\left(x^\mu, z(x^\mu)\right)$,
 we find
\begin{equation}
g^{(\rm ind)}_{\mu\nu}=\frac{1}{\chi(z)^2}
\left( g_{\mu\nu}
  -z_{,\mu}z_{,\nu}\right).
 \label{eq2002}
\end{equation}
A straightforward calculation of the determinant yields  
the brane  action in the form 
 \begin{equation}
S_{\rm br} 
=-\sigma\int d^4 x\sqrt{- g}\:
\chi^{-4}
 \sqrt{(1-X)} ,
 \label{eq0006}
\end{equation}
where we have introduced the abbreviation
\begin{equation}
X=g^{\mu\nu} z_{,\mu} z_{,\nu}. 
\end{equation}
Hence, we have obtained a 
tachyon Lagrangian with potential 
\begin{equation}
V(z)=\sigma\chi(z)^{-4},
\label{eq1007}
\end{equation}
where the fifth conformal coordinate $z=z(x)$  
has  become a dynamical tachyon field.
An attempt was made to generalize the action (\ref{eq0006}) by replacing 
$(1-X)^{1/2}$ by $(1-X)^q$ with $q$ being an arbitrary positive power \cite{choudhury}.
However, only the action with $q=1/2$  stems from $d$-brane dynamics
in a $d+1+1$ bulk.

Next,  we derive the tachyon field equations 
from  the  action (\ref{eq0006}).
The tachyon Lagrangian takes the form 
\begin{equation}
{\cal{L}} =  
-\frac{\sigma}{\chi(z)^4}\sqrt{1-g^{\mu\nu}z_{,\mu}z_{,\nu}}  \,,
 \label{eq000}
\end{equation}
and in the following we assume that the function $\chi(z)$ is known.
Note that for a pure AdS$_5$ bulk 
we have $\chi=kz$
 and we reproduce the brane action of  Ref. \cite{bilic}
if we identify the scale $k$ with the inverse of the AdS$_5$ curvature radius $\ell$.

It is important to stress that the cosmology in section \ref{braneworld}  was derived assuming
that the observer brane is moving in the  bulk with time independent geometry, 
and Eq.\ (\ref{eq027}) relates the position of  
the observer brane to the cosmological scale $a$. 
However, in this section 
we work in the gauge where the observer brane is at a fixed position and the  cosmology 
will reflect the time dependence of the  bulk metric in addition to
the time dependence of the dynamical brane position.
We consider a spatially flat FRW spacetime on the observer brane
with four dimensional line element in the standard form 
\begin{equation}
 ds^2=g_{\mu\nu}dx^\mu dx^\nu=dt^2-a(t)^2(dr^2+r^2 d\Omega^2) ,
 \label{eq0012}
\end{equation}
where, unlike in Sec.\ \ref{braneworld}, the time $t$ is synchronous.
In the cosmological context it is natural to assume that the
tachyon condensate is comoving, i.e., the velocity components are $u_\mu=(1,0,0,0)$
and $X$ 
  becomes simply
\begin{equation}
 X=\dot{z}^2 .
 \label{eq2036}
\end{equation}

The treatment of our system is conveniently performed in the
covariant Hamiltonian formalism 
based on earlier works on symplectic formalism of De Donder \cite{dedonder} and Weyl \cite{weyl}
(for recent reviews see Refs. \cite{struckmeier,cremaschini}; 
for details and application in cosmology see Ref.\ \cite{bilic}).
For this purpose we first define 
the conjugate momentum field as
\begin{equation}
\pi_z^\mu=
\frac{\partial{\cal{L}}}{\partial z_{,\mu}}.
\end{equation}
In the cosmological context  $\pi_z^\mu$ is time-like so we may also define 
its magnitude as 
\begin{equation}
\pi_z=\sqrt{g_{\mu\nu}\pi_z^\mu \pi_z^\nu}.
\label{eq2118}
\end{equation}
The Hamiltonian density may  be derived from the stress tensor corresponding to the
Lagrangian (\ref{eq000}) or by the Legendre transformation.
Either way one finds 
\begin{equation}
{\cal{H}} =\frac{\sigma}{\chi^4}\sqrt{1+\pi_{z}^2\chi^8/\sigma^2}.
 \label{eq001}
\end{equation}
Then, we can write Hamilton's equations in the form \cite{bilic}
\begin{eqnarray}
\dot{z} = \frac{\partial{\cal{H}}}{\partial\pi_z},\label{eqHam2} \\
\dot{\pi}_z + 3H\pi_z=-\frac{\partial{\cal{H}}}{\partial z}.\label{eqHam4}
\end{eqnarray}
In the spatially flat cosmology 
the Hubble expansion rate $H$ is related to the Hamiltonian via the
modified Friedmann equation (\ref{eq022}). 
As the cosmological scale is no longer related to the observer's brane position,
the function $\chi(a)$ need not satisfy the condition (\ref{eq027}).
Nevertheless, the brane  cosmology is governed by the same 
(approximate) Friedmann equation (\ref{eq005})
in which the  scale dependent gravitational constant will have a
functional dependence on the  the warp factor as dictated by equation (\ref{eq335}), i.e., $G\propto \chi_{,z}$.
However, the functional dependence on the cosmological scale 
will be subject to the field equations (\ref{eqHam2}) and (\ref{eqHam4})
together with Eq.\ (\ref{eq005}) 
which can be written as
\begin{equation} 
 H\equiv\frac{\dot{a}}{a}=\sqrt{\frac{8 \pi G_{\rm N}}{3} \mathcal{H}\left(\frac{\chi_{,z}}{k}
+ \frac{2 \pi G_{\rm N}}{3k^2} \mathcal{H}\right) }.
\label{scale_a}
\end{equation}

To solve the system of equations (\ref{eqHam2})-(\ref{scale_a}) it is convenient to
rescale the time as $t=\tau/k$ and  express  
the system in terms of dimensionless quantities. 
Besides,
by appropriately rescaling
the tachyon field $z$  and its
conjugate field $\pi_{z}$ we can eliminate  the brane tension $\sigma$ from the equations. 
To this end  we  introduce  the dimensionless functions 
\begin{eqnarray}
h = H/k, 
\quad
\varphi=k z, \quad
\pi_\varphi = \pi_z/\sigma ,
\label{eq002}
\end{eqnarray}
and we rescale the Lagrangian and Hamiltonian to obtain the 
rescaled dimensionless
pressure and energy density:
\begin{equation}
 p= \frac{\mathcal{L}}{\sigma}=-\frac{1}{\chi^4\sqrt{1+\chi^8\pi_{\varphi}^2}}=
 -\frac{1}{\chi^4}\sqrt{1-\dot{\varphi}^2} ,
 \label{eq0081}
\end{equation}
\begin{equation}
 \rho= \frac{\mathcal{H}}{\sigma}=\frac{1}{\chi^4}\sqrt{1+\chi^8\pi_{\varphi}^2}=
 \frac{1}{\chi^4}\frac{1}{\sqrt{1-\dot{\varphi}^2}}.
 \label{eq008}
\end{equation}
In these equations  and from now on the overdot denotes a derivative with respect to $\tau$.
Then we introduce a combined dimensionless coupling
\begin{equation}
\kappa^2=\frac{8\pi G_5}{k}\sigma=\frac{8\pi G_{\rm N}}{k^2}\sigma 
\label{eq102}
\end{equation}
and from (\ref{eqHam2})-(\ref{scale_a})
we obtain the following set of equations 
\begin{equation}
\dot \varphi=\frac{\chi^4\pi_{\varphi}}
{\sqrt{1+\chi^8\pi_{\varphi}^2}}
=\frac{\pi_{\varphi}}{\rho} ,
\label{eq003}
\end{equation}
\begin{equation}
\dot \pi_\varphi=-3h\pi_\varphi
+\frac{4\chi_{,\varphi}}{\chi^5
\sqrt{1+\chi^8\pi_\varphi^2}}.
\label{eq004}
\end{equation}
Here
\begin{eqnarray}
\label{h}
h=\sqrt{\frac{\kappa^2}{3}\rho\left(\chi_{,\varphi}+\frac{\kappa^2}{12}\rho  \right)} ,
\label{eq4305}
\end{eqnarray}
where 
$\chi_{,\varphi}$  
 is an abbreviation for $\partial\chi/\partial\varphi$.
Obviously, the explicit dependence on $\sigma$ and $k$ in Eqs.\ (\ref{eq003})-(\ref{eq4305}) is eliminated 
leaving one dimensionless
 free parameter $\kappa$ which could, in principle, be fixed from phenomenology.

%

%

\section{Cosmological solutions to the field equations}
\label{solutions}

Next, we analyze in detail the tachyon with potential
\begin{equation}
V(\varphi)=\varphi^{-n}, \quad \mbox{or} \quad
\chi(\varphi)=\varphi^{n/4},
\label{eq1006}
\end{equation}
As shown  in Appendix, this inverse power-law dependence for $n>4$
can be derived from the exponential superpotential (\ref{eq307})
in the braneworld model with a scalar in the bulk.
According to (\ref{eq007}) and (\ref{eq1007}), 
the potential (\ref{eq1006}) being a monotonously decreasing function of $\varphi$,
is consistent with the positivity requirement for  $G(a)$.
We will assume that the bounds on variations of $G$ with $a$ discussed in Sec.\ \ref{braneworld} 
do not apply to the pre-BBN cosmology, in particular during inflationary epoch,
so we will ignore the constraint (\ref{eq1004}). 

With (\ref{eq1006}) equations (\ref{eq003}) and (\ref{eq004}) become 
\begin{equation}
\dot \varphi=\frac{\varphi^{n}\pi_\varphi}{\sqrt{1+\pi_\varphi^2\varphi^{2n}}},
\label{eq003a}
\end{equation}
\begin{equation}
\dot \pi_\varphi=-3h\pi_\varphi
+\frac{n}{\varphi^{n+1}\sqrt{1+\pi_\varphi^2\varphi^{2n}}}.
\label{eq004a}
\end{equation}
We will try to solve these equations by the ansatz
\begin{equation}
\pi_\varphi=c_n\varphi^{m-n},
\label{eq4001}
\end{equation}
where the constants $c_n$ and   $m$ are to be fixed by the field equations.
With this ansatz the density is given by
\begin{equation}
\rho=\varphi^{-n}(1+\pi_\varphi^2\varphi^{2n})^{1/2}=\varphi^{-n}(1+c_n^2\varphi^{2m})^{1/2}
\label{eq401}
\end{equation}
and 
Eq.\ (\ref{eq003a}) becomes
\begin{equation}
\dot{\varphi}=\frac{c_n\varphi^m}{\sqrt{1+c_n^2\varphi^{2m}}} .
\label{eq4003}
\end{equation}
Furthermore, the time derivative of (\ref{eq4001}) together with (\ref{eq4003})
yields
\begin{equation}
\dot{\pi_\varphi}=c_n^2(m-n)\varphi^{2m-n-1}\left(1+c_n^2\varphi^{2m}\right)^{-1/2}.
\label{generalpi}
\end{equation}
We will look for solutions in the low and high energy density
regimes of BWC characterized by the conditions
$\kappa^2\rho/12\ll\chi_{,\varphi}$ and $\kappa^2\rho/12\gg\chi_{,\varphi}$,
respectively.
In these regimes the Hubble rate is respectively given by
\begin{equation}
h=\frac{\kappa}{\sqrt{3}}\rho^{1/2}\chi_{,\varphi}^{1/2}, \quad\quad
h=\frac{\kappa^2}{6}\rho.
\label{h1}
\end{equation}
It is advantageous to analyze  these two particular cosmologies by
making use of a  more general equation
\begin{equation}
h=h_0 \rho^\alpha \chi_{,\varphi}^\beta ,
\label{eq1001}
\end{equation}
where the constants $h_0$ and $\alpha$ are positive and $\beta<1$.
In particular,  $h_0$, $\alpha$, and $\beta$,   are 
respectively equal to $\kappa^2/6$,  1, and 0, at high density and $\kappa/\sqrt3$,
 $1/2$, and $1/2$
at low density. Equation (\ref{eq1001}) also includes  the standard cosmology analyzed
by Abramo and Finelli \cite{abramo} and the Gauss-Bonnet braneworld (GBB)
at high density \cite{lidsey,tsujikawa,calcagni1,calcagni2}.
In the standard cosmology  we have  
$h_0=\kappa/\sqrt3$, $\alpha=1/2$, and $\beta=0$ whereas
in the high density limit of  GBB we have \cite{lidsey,calcagni2}
$h_0=\kappa^{2/3}$, $\alpha=1/3$, and $\beta=0$.
In the latter scenario the coupling $\kappa^2$ is defined by the first equation (\ref{eq102})
where we have identified the mass scale $(4k)^2$ with the inverse of the GB coupling, i.e.,
we set $1/k =4\sqrt{\alpha_{\rm GB}}$.

Applying (\ref{eq4001})-(\ref{generalpi}) and (\ref{eq1001})
we obtain an identity 
 \begin{equation}
c_n^2(m-n)\varphi^{2m}+3 h_0 c_n\left(\frac{n}{4}\right)^\beta
\varphi^{m+1-\alpha n+\beta n/4-\beta}\left(1+c_n^2\varphi^{2m}\right)^{\alpha/2+1/2}
-n=0 ,
\label{eq97}
\end{equation}
which must hold for any $\varphi$.
It is easily seen that this identity will be satisfied if and only if $m=0$ and 
$n=n_{\rm cr}$, where the critical power $n_{\rm cr}$ depends on $\alpha$ and 
$\beta$, 
\begin{equation}
 n_{\rm cr}=\frac{1-\beta}{\alpha-\beta/4} .
 \end{equation}
In particular
$n_{\rm cr}$ equals 2 in the standard cosmology \cite{abramo},
3 in the high density GBB,
and, respectively, 4/3 and 1  
in the low and high density regimes of BWC.
For $n=n_{\rm cr}$ the derivative $\dot{\varphi}$ is simply a constant yielding
\begin{equation}
\varphi=\frac{c_{\rm cr}}{\sqrt{1+c_{\rm cr}^2}} \tau ,
\label{eq86}
\end{equation}
where, for simplicity, we have set the integration constant to  zero and
$c_{\rm cr}$ stands for  $c_{n_{\rm cr}}$.
As a consequence of (\ref{eq97}), the constant $c_{\rm cr}$ satisfies  the equation 
\begin{equation}
  c_{\rm cr}^2(1+c_{\rm cr}^2)^{\alpha-1}=\left(\frac{4^\beta n_{\rm cr}^{1-\beta}}{3 h_0}\right)^2,
  \label{eq1002}
 \end{equation}
which, in general,  cannot be solved for $c_{\rm cr}$.
However, for each of the four cases mentioned above, equation (\ref{eq1002}) becomes relatively simple
with solutions
\begin{equation}
c_{\rm cr}=\left\{
\begin{array}{ll}
 8\sqrt2/ (3\kappa)^2
\left(1+\sqrt{1+4\left(3\kappa/4\right)^4}\right)^{1/2}& \mbox{BWC, low density}\\
2/\kappa^2 &\mbox{BWC, high density}\\
2\sqrt2/(3\kappa^2)\left(1+\sqrt{1+9\kappa^4/4}\right)^{1/2} & \mbox{standard cosmology}\\
1/(3\kappa^2)\left(1+u^{1/3}+u^{-1/3}\right)& \mbox{GBB, high density} ,
\end{array}
\right.
\end{equation}
where, in the last line
\begin{equation}
 u=1+\frac{27\kappa^4}{2}+\frac{\sqrt{27} \kappa^2}{2}\sqrt{4+27\kappa^4} .
\end{equation}

From (\ref{eq0081}) and (\ref{eq008}) it follows that the equation of state is a
negative constant
\begin{equation}
 w\equiv \frac{p}{\rho}=-\frac{1}{1+c_{\rm cr}^2}
\end{equation}
and hence describes a dark energy fluid.  Note that in the strong coupling limit, i.e., $\kappa\rightarrow \infty$,
corresponding to large brane tensions, 
we have $c_{\rm cr}\rightarrow 0$ and the fluid 
approaches the cosmological constant.

From  (\ref{eq401})
and (\ref{eq86}) we obtain the density as a function of $\tau$
\begin{equation}
\rho(\tau)=\rho_0\tau^{-n_{\rm cr}},
\end{equation}
where
\begin{equation}
 \rho_0=\frac{\left(1+c_{\rm cr}^2\right)^{n_{\rm cr}/2+1/2}}{c_{\rm cr}^{n_{\rm cr}}} .
\end{equation}
Furthermore, from (\ref{h1}) we find that the cosmological scale 
behaves as  a power of $\tau$
\begin{equation}
a(\tau)=a_0\tau^q ,
\end{equation}
where
\begin{equation}
q=\frac{n_{\rm cr}\left(1+c_{\rm cr}^2\right)}{3c_{\rm cr}^2} .
\end{equation}
%
%
%
%
%
%
%
%
%
%

Finally $\rho$ can be expressed as a function of the cosmological scale as
\begin{equation}
\rho(a)=\rho_0\left(\frac{a}{a_0}\right)^{-n_{\rm cr}/q}=
\rho_0\left(\frac{a}{a_0}\right)^{-3c_{\rm cr}^2/(1+c_{\rm cr}^2)}.
\end{equation}
Obviously, in the limit $\kappa\rightarrow \infty$
we have  $c_{\rm cr} \rightarrow 0$,
and the universe approaches de Sitter.
In contrast, in the weak coupling limit $c_{\rm cr} \rightarrow \infty$,
the universe behaves as  dust.

For $n\neq n_{\rm cr}$ equation (\ref{eq97}) admits no solution.
Nevertheless, it can be solved in the asymp\-totic regimes of high and small $\varphi$.
In these regimes we distinguish two cases:
 a) $\varphi\rightarrow \infty$ and $m>0$ or $\varphi\rightarrow 0$ and $m<0$,
 and b) $\varphi\rightarrow \infty$ and $m<0$ or $\varphi\rightarrow 0$ and $m>0$.
 
\subsection*{a) $\varphi\rightarrow \infty$ and $m>0$ or $\varphi\rightarrow 0$ and $m<0$}
Keeping only the dominant terms in (\ref{eq97}) in the limit $\varphi\rightarrow\infty$ for $m>0$
or $\varphi\rightarrow 0$ for $m<0$
we find 
\begin{equation}
c_n^2(m-n)+3h_0\left(n/4\right)^\beta c_n^{\alpha+2}\varphi^{m\alpha-s}=0,
\label{eq3002}
\end{equation}
where
\begin{equation}
 s=n\left(\alpha-\frac{\beta}{4}\right)+\beta-1=\left(\frac{n}{n_{\rm cr}}-1\right)(1-\beta) .
\end{equation}
Then, to satisfy  (\ref{eq3002}) for any $\varphi$ we must have  
\begin{equation}
m=\frac{s}{\alpha} .
\label{eq3001}
\end{equation}
From this it follows $n\lessgtr n_{\rm cr}$ for $m\lessgtr 0$.
The constant $c_n$ is  given by
\begin{equation}
c_n=\left(\frac{1-\beta+n\beta/4}{3h_0\alpha}\right)^{1/\alpha}\left(\frac{4}{n}\right)^{\beta/\alpha}
\end{equation}
and we have
\begin{equation}
\rho=\pi_\varphi=c_n\varphi^{-(1-\beta+n\beta/4)/\alpha}.
\end{equation}
The equation of state
\begin{equation}
w=-(1-\dot{\varphi}^2)=-\frac{1}{1+c_n^2\varphi^{2s/\alpha}}
\end{equation}
approaches $0$ for large $\varphi$ with $n>n_{\rm cr}$  or small $\varphi$ with $n<n_{\rm cr}$. Hence,  the system 
(\ref{eq003a})-(\ref{eq004a}) 
has a dark-matter attractor at  $\dot{\varphi}^2=1$ or $\varphi=\tau$
in both limits.
The  large and small $\varphi$ limits correspond the of large and small $\tau$ limits, respectively.
In these limits we obtain the following behavior of the density and cosmological scale 
as functions of time
\begin{equation}
\rho(\tau)=c_n\tau^{-(1-\beta+n\beta/4)/\alpha},
\end{equation}
\begin{equation}
a(\tau)=a_0\tau^{(1-\beta+n\beta/4)/(3\alpha)},
\end{equation}
so that the density
as a function of the cosmological scale  
\begin{equation}
\rho(a)=c_n\frac{a_0^3}{a^3}
\end{equation}
clearly demonstrates  the dust behavior.

\subsection*{b) $\varphi\rightarrow \infty$ and $m<0$ or $\varphi\rightarrow 0$ and $m>0$}
This case is relevant for an inflationary scenario.
Namely, as we shall shortly see, 
the slow roll condition $\dot{\varphi} \ll 1$  for the tachyon inflation
\cite{steer,bilic3} is met in both $\varphi\rightarrow \infty$ and $\varphi\rightarrow 0$
limits.
 Keeping  the dominant terms in (\ref{eq97}) in the limit  $\varphi\rightarrow\infty$ for $m<0$
or $\varphi\rightarrow 0$ for $m>0$
 the equation
reduces to
\begin{equation}
3h_0\left(n/4\right)^\beta c_n
\varphi^{m-s}=n
\end{equation}
yielding 
\begin{equation}
m=s\equiv\left(\frac{n}{n_{\rm cr}}-1\right)(1-\beta) ,
\end{equation}
so $m\lessgtr 0$ implies $n\lessgtr n_{\rm cr}$ as before.
The coefficient $c_n$ is now given by
\begin{equation}
c_n=\frac{4^\beta n^{1-\beta}}{3h_0}.
\end{equation}
Then, using 
\begin{equation}
\rho=\varphi^{-n}, \quad\quad \pi_\varphi=c_n\varphi^{s-n} ,
\label{eq3005}
\end{equation}
we obtain
\begin{equation}
\dot{\varphi}= \pi_\varphi/\rho=c_n\varphi^s ,
\label{eq3003}
\end{equation}
from which it follows $\dot{\varphi}\rightarrow 0$ in both 
$\varphi\rightarrow \infty$ (with $s<0$) and
$\varphi\rightarrow 0$ (with $s>0$) limits. Hence,  the equation of state
$w\rightarrow-1$ in both limits.
However,  the solution to (\ref{eq3003}) critically depends on
whether $s$ is equal, greater or smaller than 1:
\begin{equation}
\varphi(\tau)=\left\{
\begin{array}{ll}
\left[c_n(1-s)(\tau-\tilde{\tau})\right]^{1/(1-s)}& \left\{
\begin{array}{l}
s<1, \tau>\tilde{\tau}\\
s>1, \tau<\tilde{\tau}
\end{array}
\right.
\\
\varphi_0\exp c_n\tau & s=1 ,
\end{array}
\right.
\label{eq3004}
\end{equation}
where  $\tilde{\tau}$ and $\varphi_0>0$ are arbitrary constants of integration.
Obviously, the limits $\varphi\rightarrow\infty$ (with $s<0$) 
and $\varphi\rightarrow 0$ (with $0< s<1$) correspond to
the limits $\tau\rightarrow\infty$ and $\tau\rightarrow \tilde{\tau}$, respectively.
The limit $\varphi\rightarrow 0$ (with $s\geq 1$) correspond to
$\tau\rightarrow -\infty$.
As a consequence of (\ref{eq3004}), the cosmological scale factor evolves as
\begin{equation}
\frac{a(\tau)}{a_0}=\left\{
\begin{array}{lcl}
\exp\left\{(\tau-\tilde{\tau})^r/\tau_0^r\right\}
& s<0, & \tau>\tilde{\tau}\\
\exp\left\{-|\tau-\tilde{\tau}|^r/\tau_0^r\right\}
&\left\{  
\begin{array}{r}
 0<s<1,\\
 s>1,
\end{array}
\right. &
\begin{array}{l}
 \tau>\tilde{\tau}\\
 \tau<\tilde{\tau}
\end{array}
\\
\exp\left\{-b_n e^{-2c_n\tau}\right\}
& s=1 ,& 
\end{array}
\right.
\label{eq3008}
\end{equation}
where 
\begin{equation}
 r=\frac{2s}{s-1}=\frac{2(n-n_{\rm cr})(1-\beta)}{(n-n_{\rm cr})(1-\beta)-n_{\rm cr}},
\end{equation}
\begin{equation}
 b_n=\frac{3h_0^2}{2\varphi_0^2 n}\left(\frac{n}{4}\right)^{2\beta} ,
\end{equation}
and
\begin{equation}
\tau_0=\left(\frac{|r|}{h_0}\right)^{1/r}
\left(\frac{4}{n}\right)^{\beta/r}
\left(c_n|s-1|\right)^{1/r-1}.
\end{equation}
Note that the limits $\varphi\rightarrow \infty$ and $\varphi\rightarrow 0$ correspond to
$a\rightarrow \infty$ and $a\rightarrow 0$, respectively.
From  (\ref{eq3005}) together with
(\ref{eq3004}) 
using the inverted relations (\ref{eq3008}) we find the density
as a function of the cosmological scale:
\begin{equation}
\rho(a)=\left\{
\begin{array}{ll}
\left[c_n|s-1|\tau_0\right]^{n/(s-1)}\left|\ln(a/a_0)\right|^{n/(2s)}&
\left\{
\begin{array}{ll}
 s<0, & a>a_0 \\
 s>0, \neq 1, &a<a_0 
\end{array}
\right. \\
\left(\varphi_0^2b_n\right)^{-n/2}\left|\ln(a/a_0)\right|^{n/2}&
s=1, \quad a<a_0. 
\end{array}
\right.
\end{equation}
Hence, in the asymptotic regimes of small and large $a$ the density varies logarithmically,
thus demonstrating a quasi de Sitter behavior.

\section{Conclusions}
\label{conclude}
We have studied a braneworld cosmology (BWC) scenario based on 
the second RS model
extended to more general warp factors. 
We have shown how our BWC is related to the cosmology of the braneworld in
the bulk with a selfinteracting scalar field minimally coupled to gravity.
In the high density regime of our BWC the modified Friedmann equation is identical to that of the original RSII
cosmology
and can be relevant for the early stages of inflation.
Within a reasonable approximation 
in the low density regime the modified Friedmann equation 
remains of the same form as in the standard cosmology  
except that the effective gravitational constant $G$ is scale dependent.

As an application we have investigated a class of tachyon models in the framework of BWC.
Assuming no restrictions on variations of $G$ in a pre-BBN cosmology we have analyzed
a power-law variation 
corresponding to the inverse power-law tachyon potential $V\propto \varphi^{-n}$.
We have demonstrated a universal critical behavior for the cosmologies described by (\ref{eq1001}) 
for the tachyon field theory with inverse power-law potential:
there exist a critical power $n=n_{\rm cr}$ that divides
a dust universe  for $n>n_{\rm cr}$ and a quasi de Sitter universe for 
$0<n<n_{\rm cr}$ in both  asymptotic regimes of large and small tachyon field $\varphi$
with $n_{\rm cr}$  depending on the details of the cosmological scenarios. 
In particular we have analyzed three different scenarios: 
the standard tachyon cosmology and low and high energy-density regimes of the 
braneworld cosmology. For these three cosmologies, we have found $n_{\rm cr}$ to be equal to
2, 4/3, and 1, respectively.

\section*{Acknowledgments}

This work has been supported by the Croatian 
Science Foundation under the project IP-2014-09-9582
and partially supported by 
ICTP - SEENET-MTP project PRJ-09 Cosmology and Strings.
The work of N.\ Bili\'c and S.\ Domazet 
has been partially supported by the H2020 CSA Twinning project No.\ 692194, “RBI-T-WINNING”.
G.\ Djordjevic  acknowledges support by the Serbian Ministry for Education, 
Science and Technological Development under the projects No.\ 176021 
and No. 174020. 

 \appendix

\section{Scalar field in the bulk}
\label{scalar}
 
Following DeWolfe et al.\ \cite{dewolfe} we generalize the original RSII action
to include  a minimally coupled scalar field $\Phi$
in the bulk
so that in the absence of $\Phi$ the bulk is reduced to the pure AdS$_5$.
The dynamics of a 3-brane in a 4+1 dimensional bulk is described by 
the total action as the sum of the bulk and  brane actions  
\begin{equation}
 S=S_{\rm bulk}+S_{\rm br} .
 \label{eq4002}
\end{equation}
The bulk action is given by
\begin{equation} 
S_{\rm bulk} =\frac{1}{8\pi G_5}\int d^5X \sqrt{G} 
\left[-\frac{R^{(5)} }{2} +\frac12 G^{ab}\Phi_{,a} \Phi_{,b} -U(\Phi)\right] 
+S_{\rm GH}[h],
\end{equation}
where 
the Gibbons-Hawking boundary term  is
  given by an integral over the brane hypersurface $\Sigma$:
\begin{equation} 
S_{\rm GH}[h] =\frac{1}{8\pi G_5}\int_\Sigma d^{4}x\sqrt{-h} K[h] .
\label{eq203}
\end{equation} 
The quantity $K$ is the trace of the extrinsic curvature tensor $K_{ab}$
defined as 
\begin{equation}
K_{ab}=h_{a}^{c}h_{b}^{d} n_{d;c} ,
\label{eq109}
\end{equation}
where $n^a$ is a unit vector normal to the brane
pointing towards increasing fifth coordinate,
$h_{ab}$ is 
 the induced metric
\begin{equation}
 h_{ab}=G_{ab}+n_a n_b ,
 \label{eq1003}
\end{equation}
and $h\equiv \det h_{ab}$ is its determinant.
Observers  on the brane with action
\begin{equation} 
S_{\rm br}[h] =-\int_\Sigma d^{4}x\sqrt{-h} \sigma(\Phi),
\label{eq1005}
\end{equation}
see the induced metric $h_{\mu\nu}$.
As we will shortly see, the relationship between the brane tension $\sigma$ and 
the potential $U$ 
is dictated by the field equations and the boundary conditions
on the brane.

Next we outline a derivation of
 the RSII like solution assuming  
the metric and bulk scalar field  depend on the fifth coordinate only
\cite{dewolfe}.
The Einstein equations in the bulk are
\begin{equation}
 6\frac{\psi^{\prime 2}}{\psi^2}=\frac12 \Phi^{\prime 2} -U(\Phi) ,
 \label{eq4004}
\end{equation}
\begin{equation}
 -3\frac{\psi''}{\psi}+3\frac{\psi^{\prime 2}}{\psi^2}= \Phi^{\prime 2}, 
 \label{eq4005}
\end{equation}
together with the $\Phi$ field equation away from the brane 
\begin{equation}
 \Phi'' +4 \frac{\psi'}{\psi}\Phi'= \frac{dU}{d\Phi},
 \label{eq4006}
\end{equation}
where the prime $'$ denotes a derivative with respect to $y$.
In addition, the solutions are subject to the junction conditions on the brane
\begin{equation}
 - 3\left[ \left[\frac{\psi^{\prime}}{\psi}\right]\right] =8\pi G_5\sigma(\Phi_0),
\label{eq303}
\end{equation}
\begin{equation}
\left[ \left[\Phi'\right]\right] = 8\pi G_5\frac{\partial \sigma(\Phi_0)}{d\Phi_0},
\label{eq304}
\end{equation}
where $[[f]]$ denotes the discontinuity of a function $f(y)$ across the brane,
i.e.,
\begin{equation}
\left[ \left[ f (y) \right] \right] =
\lim_{\epsilon \rightarrow 0}  \,
\left( f (y_{\rm br} + \epsilon) - f (y_{\rm br}-\epsilon) \right) .
\end{equation}  
Eq.\ (\ref{eq303}) is a generalized  fine tuning condition of RSII.

The system of equations (\ref{eq4004})-(\ref{eq4006}) can be reduced to three
decoupled first order differential equations \cite{dewolfe}.
Suppose $U(\Phi)$ is expressed in terms of a superpotential $W(\Phi)$
\begin{equation}
U(\Phi)=\frac18 \left( \frac{dW}{d\Phi}  \right)^2 -\frac16 \left(W(\Phi)\right)^2.
 \label{eq3007}
\end{equation}
Then, a solution to equations
\begin{equation}
 \frac{\psi'}{\psi} =\mp\frac16 W(\Phi), \quad \quad \Phi'=\pm \frac12 \frac{dW}{d\Phi},
 \label{eq4007}
\end{equation}
will also satisfy equations (\ref{eq4004})-(\ref{eq4006}). 

Owing to the assumed  orbifold symmetry, 
the superpotential $W$ and its derivative $dW/d\Phi$ will  be  continuous across
the brane if the upper and lower signs in (\ref{eq4007}) are chosen for $y>y_{\rm br}$ and $y<y_{\rm br}$,
respectively.
Then, the junction conditions (\ref{eq303}) and (\ref{eq304}) will be satisfied provided
\begin{equation}
\sigma(\Phi_0)=\frac{1}{8\pi G_5} W(\Phi_0), \quad \frac{\partial \sigma(\Phi_0)}{d\Phi_0}=
\frac{1}{8\pi G_5} \frac{dW (\Phi_0)}{d\Phi_0} ,
 \label{eq302}
\end{equation}
where $\Phi_0$ denotes the value of $\Phi$ on the brane.
Hence, generally $\sigma(\Phi)$ is tangent to $W(\Phi)$ on the brane.
A stronger condition $\sigma(\Phi)\equiv  W(\Phi)/(8\pi G_5)$ 
is required for a BPS brane\cite{kim2}.

The system of equations (\ref{eq4007}) can be expressed in terms of $\chi$ as a single second order
differential equation. In the region $y_{\rm br}\leq y<\infty$ ($z_{\rm br}\leq z<\infty$) one finds 
\begin{equation}
\chi\frac{d^2\chi}{dz^2} - \frac{1}{12} \left( \frac{dW}{d\Phi}\right)^2
=0,
 \label{eq305}
\end{equation}
where the argument of  $dW/d\Phi$  is 
\begin{equation}
 \Phi=W^{-1}(6\chi_{,z}).
 \label{eq306}
\end{equation}
Here $W^{-1}$ denotes the inverse  function  of $W$ 
and $\chi_{,z}$ is an abbreviation for $d\chi/dz$.
Similarly, in the region $-\infty<y\leq y_{\rm br}$ ($0<z\leq z_{\rm br}$) one finds
\begin{equation}
\frac{z^2}{z_{\rm br}^2}\chi\frac{d}{dz}\left(\frac{z^2}{z_{\rm br}^2}\frac{d\chi}{dz}\right)
- \frac{1}{12} \left( \frac{dW}{d\Phi}\right)^2
=0 ,
 \label{eq311}
\end{equation}
where the argument of  $dW/d\Phi$ is 
\begin{equation}
 \Phi=W^{-1}(-6z^2\chi_{,z}/z_{\rm br}^2) .
 \label{eq312}
\end{equation}

As an example of particular interest, consider  a warp function of the power-law form
 \begin{equation}
\chi(z)=\left\{ 
\begin{array}{ll}
(z/z_{\rm br})^n ,& \mbox{for $z_{\rm br}\leq z<\infty$}\\ 
(z_{\rm br}/z)^n, & \mbox{for $0<z\leq z_{\rm br}$}
\end{array}
\right.
 \label{eq309}
\end{equation}
Then, using either (\ref{eq305}) with (\ref{eq306}) in the region $z_{\rm br}\leq z<\infty$
or (\ref{eq311}) with  (\ref{eq312}) in the region $0<z\leq z_{\rm br}$,
we obtain 
the differential equation 
\begin{equation}
\left(\frac{dW}{d\Phi}\right)^2-\eta W^2=0,
 \label{eq308}
\end{equation}
where
\begin{equation}
 \eta=\frac{n-1}{3n}.
\end{equation}
For $n\geq 1$,  equation (\ref{eq308}) has a real solution 
\begin{equation}
 W(\Phi)= 6k e^{\sqrt{\eta}\Phi}.
 \label{eq307}
\end{equation}
In the limiting case of $n=1$ ($\eta=0$) one recovers 
the RSII model with
pure AdS$_5$ bulk  with curvature radius $\ell= 1/k=z_{\rm br}$.
The superpotentials of the exponential form (\ref{eq307}) have been extensively
studied in the context of braneworlds inspired by supergravity \cite{brax2,davis,brax1,brax3}.

For $n<1$, Eq.\ (\ref{eq308}) has no real solution. 
However, the power-law warp factor of the form
\begin{equation}
\chi=\left(\frac{z}{z_{\rm br}}\right)^n-C ,
 \label{eq321}
\end{equation}
with $C>0$, yields a modified differential equation
\begin{equation}
\left(\frac{dW}{d\Phi}\right)^2-\eta W^2\left(1-C\left(\frac{z_{\rm br}W}{6n}\right)^{-1/(3\eta)} \right)=0,
 \label{eq322}
\end{equation}
which 
has real solutions for any real $n$.
The solution can be expressed in the form
 \begin{equation}
W=\frac{6nC^{3\eta}}{z_{\rm br}}\times\left\{ 
\begin{array}{ll}
 \left[1+\tan^2(\Phi/(6\sqrt{-\eta})+C_1)\right]^{-3\eta},& \mbox{for $n< 1$}\\ 
\left[1-\tanh^2(\Phi/(6\sqrt{\eta})+C_2)\right]^{-3\eta}, & \mbox{for $n\geq 1$},
\end{array}
\right.
 \label{eq323}
\end{equation}
where $C_1$ and $C_2$ are arbitrary integration constants
which can be conveniently chosen. For example, with the choice
\begin{equation}
C_2=\frac{1}{6\eta} \ln \frac{4^{3\eta}}{n C^{3\eta}}
 \label{eq324}
\end{equation}
one recovers the solution (\ref{eq307}) in the limit $C\rightarrow 0$.


\begin{thebibliography}{1}
\bibitem{arkani}
N.~Arkani-Hamed, S.~Dimopoulos, and G.~Dvali,
Phys.\ Lett.\ B {\bf 429}, 263 (1998).
\bibitem{antoniadis} 
  I.~Antoniadis, N.~Arkani-Hamed, S.~Dimopoulos and G.~R.~Dvali,
  Phys.\ Lett.\ B {\bf 436}, 257 (1998)
  [hep-ph/9804398].

%
\bibitem{randall1}
L.~Randal and R.~Sundrum, Phys.\ Rev.\ Lett. {\bf 83}, 3370 (1999)
\bibitem{randall2}
L.~Randal and R.~Sundrum, Phys.\ Rev.\ Lett. {\bf 83}, 4690 (1999)
%
\bibitem{gubser2} 
  S.~S.~Gubser,
  Phys.\ Rev.\ D {\bf 63}, 084017 (2001)
  [hep-th/9912001];
%
  S.~Nojiri, S.~D.~Odintsov and S.~Zerbini,
  Phys.\ Rev.\ D {\bf 62}, 064006 (2000)
  [hep-th/0001192];
%
  S.~B.~Giddings, E.~Katz and L.~Randall,
  JHEP {\bf 0003}, 023 (2000)
  [hep-th/0002091];
%
  S.~W.~Hawking, T.~Hertog and H.~S.~Reall,
  Phys.\ Rev.\ D {\bf 62}, 043501 (2000)
  [hep-th/0003052];
%
  M.~J.~Duff and J.~T.~Liu,
  Class.\ Quant.\ Grav.\  {\bf 18}, 3207 (2001)
  [Phys.\ Rev.\ Lett.\  {\bf 85}, 2052 (2000)]
  [hep-th/0003237];
 %
  S.~Nojiri and S.~D.~Odintsov,
  Phys.\ Lett.\ B {\bf 484}, 119 (2000)
  [hep-th/0004097];
%
  S.~de Haro, K.~Skenderis and S.~N.~Solodukhin,
  Class.\ Quant.\ Grav.\  {\bf 18}, 3171 (2001)
  [hep-th/0011230]; 
%
\bibitem{bilic1} 
  N.~Bilic,
  Phys.\ Rev.\ D {\bf 93}, no. 6, 066010 (2016)
  [arXiv:1511.07323 [gr-qc]]. 
%
%
  \bibitem{binetruy} 
  P.~Binetruy, C.~Deffayet and D.~Langlois,
  Nucl.\ Phys.\ B {\bf 565}, 269 (2000)
  [hep-th/9905012]. 
  \bibitem{flanagan} 
  E.~E.~Flanagan, S.~H.~Henry~Tye and I.~Wasserman,
  Phys.\ Rev.\ D {\bf 62}, 044039 (2000)
  [hep-ph/9910498].
   \bibitem{godlowski} 
  W.~Godlowski and M.~Szydlowski,
  Gen.\ Rel.\ Grav.\  {\bf 36}, 767 (2004)
  [astro-ph/0404299]. 
 %
\bibitem{goldberger} 
  W.~D.~Goldberger and M.~B.~Wise,
  Phys.\ Rev.\ Lett.\  {\bf 83}, 4922 (1999)
  [hep-ph/9907447].
%
  \bibitem{kobayashi} 
  S.~Kobayashi, K.~Koyama and J.~Soda,
  Phys.\ Rev.\ D {\bf 65}, 064014 (2002)
  [arXiv:hep-th/0107025].
%
%
    \bibitem{dewolfe} 
 O.~DeWolfe, D.~Z.~Freedman, S.~S.~Gubser and A.~Karch,
 Phys.\ Rev.\ D {\bf 62}, 046008 (2000)
 [hep-th/9909134].
 %
  
 \bibitem{german1}
  G.~German, A.~Herrera-Aguilar, D.~Malagon-Morejon, R.~R.~Mora-Luna and R.~da Rocha,
  JCAP {\bf 1302}, 035 (2013)
  [arXiv:1210.0721 [hep-th]].
%
 \bibitem{german2}
   G.~Germán, A.~Herrera-Aguilar, A.~M.~Kuerten, D.~Malagon-Morejon and R.~da Rocha,
  JCAP {\bf 1601}, no. 01, 047 (2016)
  [arXiv:1508.03867 [hep-th]].
%
 \bibitem{Bogdanos:2006dt}
 C.~Bogdanos, A.~Dimitriadis and K.~Tamvakis,
 Class.\ Quant.\ Grav.\  {\bf 24} (2007) 3701
 doi:10.1088/0264-9381/24/14/010
 [hep-th/0611181].
 
 \bibitem{HerreraAguilar:2011jm}
 A.~Herrera-Aguilar, D.~Malagon-Morejon, R.~R.~Mora-Luna and I.~Quiros,
 Class.\ Quant.\ Grav.\  {\bf 29} (2012) 035012
 doi:10.1088/0264-9381/29/3/035012
 [arXiv:1105.5479 [hep-th]].
 
  \bibitem{sen} 
  A.~Sen,
  JHEP {\bf 9910}, 008 (1999)
  [hep-th/9909062].
  \bibitem{shandera} 
  S.~E.~Shandera and S.-H.~H.~Tye,
  JCAP {\bf 0605}, 007 (2006)
  [hep-th/0601099]. 
  \bibitem{abramo}
  L.~R.~W.~Abramo and F.~Finelli,
  Phys.\ Lett.\ B {\bf 575}, 165 (2003)
  [astro-ph/0307208]. 
    \bibitem{kofman2} 
  L.~Kofman, A.~D.~Linde and A.~A.~Starobinsky,
  Phys.\ Rev.\ Lett.\  {\bf 73}, 3195 (1994)
  [hep-th/9405187].
  \bibitem{kofman3} 
  L.~Kofman, A.~D.~Linde and A.~A.~Starobinsky,
  Phys.\ Rev.\ D {\bf 56}, 3258 (1997)
  [hep-ph/9704452].
  \bibitem{bassett} 
  B.~A.~Bassett, S.~Tsujikawa and D.~Wands,
  Rev.\ Mod.\ Phys.\  {\bf 78}, 537 (2006)
  [astro-ph/0507632]. 
  %
  
  \bibitem{bilic}
  N.~Bili\'c and G.~B.~Tupper,
  Central Eur.\ J.\ Phys.\  {\bf 12}, 147 (2014)
  [arXiv:1309.6588 [hep-th]].
  
  \bibitem{israel}
  W. Israel, {\it Nuovo Cim.} B {\bf 44} S10, 1 (1966); Erratum-ibid B 
  {\bf 48}, 463 (1967).
  
   \bibitem{brax2} 
  P.~Brax and C.~van de Bruck,
  Class.\ Quant.\ Grav.\  {\bf 20}, R201 (2003)
  [hep-th/0303095]; 
  P.~Brax, C.~van de Bruck and A.~C.~Davis,
  Rept.\ Prog.\ Phys.\  {\bf 67}, 2183 (2004)
  [hep-th/0404011].
\bibitem{bernardini} 
  A.~E.~Bernardini, R.~T.~Cavalcanti and R.~da Rocha,
  Gen.\ Rel.\ Grav.\  {\bf 47}, no. 1, 1840 (2015)
  doi:10.1007/s10714-014-1840-x
  [arXiv:1411.3552 [gr-qc]].  
  \bibitem{accetta}
  F.~S.~Accetta, L.~M.~Krauss and P.~Romanelli,
  Phys.\ Lett.\ B {\bf 248}, 146 (1990).
  \bibitem{uzan} 
  J.~P.~Uzan,
  Rev.\ Mod.\ Phys.\  {\bf 75}, 403 (2003)
  [hep-ph/0205340]; 
  Living Rev.\ Rel.\  {\bf 14}, 2 (2011)
  [arXiv:1009.5514 [astro-ph.CO]].
  %
  \bibitem{planck2015} 
  P.~A.~R.~Ade {\it et al.} [Planck Collaboration],
  Astron.\ Astrophys.\  {\bf 594}, A13 (2016)
  [arXiv:1502.01589 [astro-ph.CO]]. 
  %
    \bibitem{amarilla} 
  L.~Amarilla and H.~Vucetich,
  Int.\ J.\ Mod.\ Phys.\ A {\bf 25}, 3835 (2010)
  [arXiv:0908.2949 [gr-qc]].  
  %
  
  \bibitem{davis} 
  A.~C.~Davis, P.~Brax and C.~van de Bruck,
  Nucl.\ Phys.\ Proc.\ Suppl.\  {\bf 148}, 64 (2005)
  [astro-ph/0503467].
  %
  
  \bibitem{bertotti}
  B.~Bertotti, L.~Less, and P.~Tortora, Nature {\bf 425}, 374 (2003).
  %

  
  \bibitem{lykken} 
  J.~D.~Lykken and L.~Randall,
  JHEP {\bf 0006}, 014 (2000)
  [hep-th/9908076].
  
  \bibitem{bordemann}
  M.~Bordemann and J.~Hoppe,
  Phys.\ Lett.\ B {\bf 325}, 359 (1994);
  %
  N.~Ogawa,
  Phys.\ Rev.\ D {\bf 62}, 085023 (2000).
  \bibitem{jackiw}
  R.~Jackiw, (2002)
  {\it Lectures on Fluid Mechanics} (Springer Verlag, Berlin, 2002). 
  %
  \bibitem{choudhury} 
  S.~Choudhury and S.~Panda,
  Eur.\ Phys.\ J.\ C {\bf 76}, 278 (2016)
  [arXiv:1511.05734 [hep-th]].
   \bibitem{dedonder} 
  Th.~De Donder, {\it Th\'{e}orie Invariantive Du Calcul des Variations} (Gaultier-
  Villars \& Cia., Paris, 1930).  
  \bibitem{weyl} 
  H.~Weyl, Annals of Mathematics {\bf 36}, 607 (1935).
  \bibitem{struckmeier}   
  J.\ Struckmeier, A.\ Redelbach,
  Int. J. Mod. Phys. E {\bf 17} 435-491 (2008), 
  [arXiv:0811.0508 [math-ph]].
  \bibitem{cremaschini} 
  C.~Cremaschini and M.~Tessarotto,
  Appl.\ Phys.\ Res.\  {\bf 8}, 60 (2016)
  [arXiv:1609.04422 [gr-qc]].
  %
  %
  \bibitem{lidsey} 
  J.~E.~Lidsey and N.~J.~Nunes,
  Phys.\ Rev.\ D {\bf 67}, 103510 (2003)
  [astro-ph/0303168].
  \bibitem{tsujikawa} 
  S.~Tsujikawa, M.~Sami and R.~Maartens,
  Phys.\ Rev.\ D {\bf 70}, 063525 (2004)
  [astro-ph/0406078].
  \bibitem{calcagni1} 
  G.~Calcagni and S.~Tsujikawa,
  Phys.\ Rev.\ D {\bf 70}, 103514 (2004)
  [astro-ph/0407543].
  \bibitem{calcagni2} 
    G.~Calcagni,
  hep-ph/0503044.
 \bibitem{steer}
 D.~A.~Steer and F.~Vernizzi,
 Phys.\ Rev.\ D {\bf 70}, 043527 (2004)
 [hep-th/0310139].
 \bibitem{bilic3} 
 N.~Bilic, D.~Dimitrijevic, G.~Djordjevic and M.~Milosevic,
 Int.\ J.\ Mod.\ Phys.\ A {\bf 32}, 1750039 (2017)
 [arXiv:1607.04524 [gr-qc]]. 
  

\bibitem{kim2} 
  J.~E.~Kim, G.~B.~Tupper and R.~D.~Viollier,
  Phys.\ Lett.\ B {\bf 612}, 293 (2005)
  [hep-th/0503097].
  
  \bibitem{brax1} 
  P.~Brax and A.~C.~Davis,
  Phys.\ Lett.\ B {\bf 497}, 289 (2001)
  [hep-th/0011045]. 

\bibitem{brax3} 
  P.~Brax and A.~C.~Davis,
  Phys.\ Lett.\ B {\bf 513}, 156 (2001)
  [hep-th/0105269].
%




 

\end{thebibliography}
\end{document}